\documentclass{fundam}

\publyear{22}
\papernumber{2102}
\volume{185}
\issue{1}

\usepackage[T1]{fontenc}
\usepackage[utf8]{inputenc}
\usepackage{graphicx}
\usepackage{amsmath}
\usepackage{amssymb}
\usepackage{amsfonts}
\usepackage{stmaryrd}
\usepackage{tikz}
\usepackage{enumitem}
\usetikzlibrary{automata}
\usetikzlibrary{shapes,calc}
\usepackage{geometry}
\usepackage{hyperref}

\usepackage{cleveref}

\usepackage{etoolbox}
\newcommand{\FIcrefalias}[1]{\AtBeginEnvironment{#1}{\crefalias{definition}{#1}}}
\FIcrefalias{theorem}\FIcrefalias{lemma}\FIcrefalias{proposition}
\FIcrefalias{corollary}\FIcrefalias{remark}\FIcrefalias{example}
\FIcrefalias{claim}\FIcrefalias{fact}\FIcrefalias{assumption}
\FIcrefalias{conjecture}

\usepackage{todonotes}

\tikzstyle{initial}+=[initial by arrow,initial text={}]
\tikzstyle{every state}+=[minimum size=0.8cm,fill, top color=white, bottom color=black!20]

\newcommand{\NN}{\mathbb{N}}

\newcommand{\RR}{\mathbb{R}}

\newcommand{\Q}{\mathbb{Q}}

\newcommand{\<}{\langle}
\renewcommand{\>}{\rangle}
\newcommand{\lv}{\lVert}
\newcommand{\rv}{\rVert}

\newcommand{\bsigma}{\bar{\sigma}}
\newcommand{\btau}{\bar{\tau}}

\newcommand{\Circle}{\bigcirc}

\newcommand{\playcircle}{\circ}

\renewcommand{\l}{\ell}
\renewcommand{\epsilon}{\varepsilon}
\renewcommand{\phi}{\varphi}

\newcommand{\SC}{\mathsf{SC}}
\newcommand{\Plays}{\mathsf{Plays}}
\newcommand{\Hist}{\mathsf{Hist}}

\newcommand{\Conv}{\mathsf{Conv}}

\newcommand{\nego}{\mathsf{nego}}
\newcommand{\lCons}{\lambda\mathsf{Cons}}
\newcommand{\Cons}{\mathsf{Cons}}
\newcommand{\lRat}{\lambda\mathsf{Rat}}

\newcommand{\last}{\mathsf{last}}
\newcommand{\first}{\mathsf{first}}
\newcommand{\Occ}{\mathsf{Occ}}
\newcommand{\Inf}{\mathsf{Inf}}
\newcommand{\MP}{\mathsf{MP}}

\newcommand{\dseal}{\!\,^\llcorner}

\renewcommand{\Game}{\mathcal{G}}
\newcommand{\Eq}{\mathcal{E}}

\newcommand{\NP}{\mathsf{NP}}
\newcommand{\ExpTime}{\mathsf{EXPTIME}}
\newcommand{\PTime}{\mathsf{P}}

\newcommand{\PSpace}{\mathsf{PSPACE}}

\newcommand{\bx}{\bar{x}}
\newcommand{\by}{\bar{y}}
\newcommand{\bz}{\bar{z}}

\newcommand{\bbx}{\bar{\bar{x}}}
\newcommand{\bby}{\bar{\bar{y}}}

\newcommand{\bs}{\mathsf{bs}}
\newcommand{\wnego}{\mathsf{wnego}}

\crefname{algorithm}{Algorithm}{Algorithms}
\crefname{figure}{Figure}{Figures}
\crefname{lemma}{Lemma}{Lemmas}
\crefname{theorem}{Theorem}{Theorems}
\crefname{proposition}{Proposition}{Propositions}
\crefname{corollary}{Corollary}{Corollaries}
\crefname{definition}{Definition}{Definitions}
\crefname{remark}{Remark}{Remarks}
\crefname{claim}{Claim}{Claims}
\crefname{example}{Example}{Examples}
\crefname{conjecture}{Conjecture}{Conjectures}

\usepackage{url} 
\usepackage[ruled,lined]{algorithm2e}
\usepackage{graphicx}
\usepackage{tikz}
\usetikzlibrary{automata, positioning, arrows}

\begin{document}

\title{The Weak Negotiation Function}

\author{Léonard Brice\\
Institute of Science and Technology Austria \\
leonard.brice@ist.ac.at}

\maketitle

\address{leonard.brice@ist.ac.at}

\runninghead{L. Brice}{The Weak Negotiation Function}

\begin{abstract}
  Subgame-perfect equilibria (SPEs) constitute a refinement of Nash equilibria, where the players are not allowed to resort to non-credible threats to discourage deviations.
  In graph games equipped with a broad class of objectives, SPEs are captured by a tool called the negotiation function.

    After giving a quick overview of what has already been achieved using this tool, we define a variant of it: the weak negotiation function.
  We show that weak SPEs, a standard relaxation of SPEs, can be captured by the weak negotiation function exactly like SPEs are captured by the negotiation function.
  This characterisation yields a new computational result: deciding whether a given game contains a weak SPE where each player's payoff lies in a given interval is an $\NP$-complete problem if the game has mean-payoff objectives.
    The proof reuses many arguments from its SPE counterpart, demonstrating the robustness of the tool.
  
  We also use the weak negotiation function to present a new proof, significantly shorter than the known one, of the existence of weak SPEs in mean-payoff games.
  We argue that those new techniques should be reusable in other settings, in particular to prove an open conjecture about the existence of randomised $\epsilon$-SPEs in mean-payoff games.
\end{abstract}


\section{Introduction}\label{sec:intro}

\paragraph{Nash equilibria.}
A Nash equilibrium (NE) is a strategy profile from which no player can deviate profitably, given the strategies of the others.
As a fundamental solution concept in game theory, NEs are widely used to model rational behaviors of multiple agents, or to prescribe a collective behavior that comes with the guarantee that no agent is tempted to deviate from it.

\paragraph{Subgame-perfect equilibria.}
Unfortunately, in sequential games, NEs suffer from the problem of {\em non-credible threats}~\cite[Chapter 5.5]{Osborne}: some players may be discouraged to deviate only because other players threaten them with actions that are not rational, with regard to their own interests.
To avoid such dynamics, the stronger concept of {\em subgame-perfect equilibrium} (SPE) is often preferred: those are defined as profile of strategies where, after any possible history, the strategies that the players plan to follow afterwards still form an NE.
Then, SPEs ensure rationality not only in players’ effective actions, but also in their planned responses to alternative scenarios.

\paragraph{Weak NEs and weak SPEs.}
NEs and SPEs have little brothers: weak NEs, and weak SPEs.
In a weak NE, non-profitability applies only to \emph{finite} deviations, that is, when the deviating player eventually complies with the prescribed strategy.
Weak SPEs are defined as expected: after every possible history, the players' substrategies form a weak NE.
Weak SPEs have sometimes been preferred to SPEs, mostly for one property: there always exists one in every prefix-independent game~\cite{existenceWSPE}, whereas SPEs may not~\cite{SOLAN2003911}.

Another reason for that preference may be that SPEs in graph games have, for some time, resisted most algorithmic studies, despite some results existing in some specific settings~\cite{DBLP:conf/fsttcs/Ummels06,DBLP:journals/lmcs/BrihayeBGRB20}.
This has led to an intensive line of work about algorithms to compute weak SPEs in graph games~\cite{DBLP:conf/csl/BrihayeBMR15,DBLP:journals/iandc/BrihayeBGR21,DBLP:journals/ipl/Goeminne20}.

\paragraph{A tool to handle SPEs: the negotiation function.}
Later on, a new line of research defined the \emph{negotiation function}, a tool to characterise SPEs.
The negotiation function takes as input a mapping $\lambda$, which labels each vertex by a finite or infinite value, expressing the minimal payoff that the player controlling that vertex would require if that vertex is visited.
It returns a requirement $\nego(\lambda) \geq \lambda$, which labels each vertex by the best payoff that the player controlling that vertex can ensure from there if the other players play rationally with regard to the requirement $\lambda$.

If $\lambda_0$ is the \emph{vacuous requirement}, mapping every vertex to $-\infty$ (or the minimal payoff possible in the game), then $\lambda_1 = \nego(\lambda_0)$ labels every vertex by what is known in the literature as its \emph{adversarial value}, \emph{max-min value}, or simply \emph{value}.
It is well-known that, in a broad class of games,\footnote{The precise class would be that of prefix-independent games where the minimum, in the definition of the adversarial value, is always realised.} the plays that are generated by NEs are exactly the plays where each player gets at least the adversarial value of each of their vertices visited; that is, the plays that are \emph{consistent} with the requirement $\lambda_1$.
But the negotiation function enables us to define a sequence of iterates $\lambda_0$, $\lambda_1$, $\lambda_2$, \dots, which converges to a least fixed point $\lambda^*$: then, that least fixed point characterises SPEs in the same way that $\lambda_1$ characterises NEs~\cite{LMCS23}.

Such fixed point characterisations are quite common in computer science.
However, defining the negotiation function as a function that can be described analytically, instead of a mere step in an iterative process, unlocked new potentialities for the study of SPEs, which resulted in a new line of work~\cite{LMCS23,jacm,mfcs23} now gathered in the author's PhD thesis~\cite{mythesis}.

This paper is meant to demonstrate that those potentialities are not exhausted, by closing the loop: there is a variant of the negotiation function that, similarly, describes efficiently weak SPEs.

\paragraph{Contributions and structure.}
After introducing the necessary background (\Cref{sec:background}), we recall the definition of the negotiation function, its history, and its main known applications (\Cref{sec:nego}).
We then move on to weak SPEs.

In \Cref{sec:weak_nego}, we show that a variant of the negotiation function, the \emph{weak negotiation function}, characterises weak SPEs in exactly the same way as the negotiation function characterises SPEs~(\Cref{thm:wnego}).
The definition of the weak negotiation function is much simpler than that of the standard negotiation function, and the characterisation applies to a broader class of games: \emph{prefix-linear games with steady weak negotiation}.\footnote{Technically, steady negotiation does not imply steady weak negotiation, but we show in \Cref{lm:steady_steadyw} that it does so up to a minor assumption.}

In \Cref{sec:algos}, we use that characterisation to prove a new computational result: the constrained existence problem of weak SPEs is $\NP$-complete in mean-payoff games (\Cref{thm:weakSPEs_mp}).

Finally, in \Cref{sec:existence}, we use the weak negotiation function to re-prove another known result: weak SPEs always exist in mean-payoff games.
We argue that this demonstrates the flexibility of the concept, and its potential power to address new problems; as a main candidate, we cite a conjecture by Flesch, Kuipers, Schoenmakers, and Vrieze \cite[abstract]{DBLP:journals/ijgt/KuipersFSV16} about the existence of randomised $\epsilon$-SPEs in every mean-payoff game.

\section{Background}\label{sec:background}

        \subsection{Games and strategies}

In all what follows, we will use the word \emph{game} for infinite duration turn-based games with perfect information played on \emph{graphs}.

\paragraph{Graphs.}

A graph is an ordered pair $(V, E)$, where $V$ is a set of \emph{vertices} and $E \subseteq V \times V$ is a set of \emph{edges}.
For simplicity, an edge $(v, w) \in E$ will often be written $vw$.
A (finite or infinite) \emph{path} in $(V, E)$ is a word $p = p_0 p_1 \dots$ over the alphabet $V$, such that for every $k$ such that $p_k$ and $p_{k+1}$ exist, we have $p_k p_{k+1} \in E$.
Given a path $p$, we denote by $\Occ(p)$ the set of vertices that occur in $p$, and by $\Inf(p)$ the set of vertices that occur infinitely often in $p$ (which is empty if $p$ is finite).
We say that $p$ is \emph{simple} if no vertex occurs more than once.

For a given index $k$, we denote by $p_{\leq k}$, or $p_{< k+1}$, the finite prefix $p_0 \dots p_k$, and $p_{\geq k}$, or $p_{> k-1}$, the (finite or infinite) suffix $p_k p_{k+1} \dots$.
Finally, we write $\first(p)$ for the first vertex of $p$, and (if $p$ is finite) $\last(p)$ for the last vertex of $p$.

Given a finite path $p = p_0 \dots p_{n-1}$, its \emph{length} is $|p| = n$.
It is a \emph{cycle} if we have $p_{n-1} p_0 \in E$.
A \emph{lasso} is a path of the form $hc^\omega$, where $c$ is a cycle.

\paragraph{Games.}
We have two notions of games, with and without an initial vertex.

\begin{definition}[Non-initialised game]\label{defi_game}
	A \emph{non-initialised game} is a tuple
	$\Game = \left(\Pi, V, (V_i)_{i \in \Pi}, E, \mu\right)$, where:
	
	\begin{itemize}
		\item $\Pi$ is a finite set of \emph{players};
		
		\item $(V, E)$ is a directed graph, called the \emph{underlying graph} of $\Game$, in which every vertex has at least one outgoing edge;

		\item $(V_i)_{i \in \Pi}$ is a partition of $V$, in which $V_i$ is the set of vertices \emph{controlled} by player $i$;
		
		\item $\mu$ is a \emph{payoff function}, that maps each path $\pi$ in $(V, E)$ to the tuple $\mu(\pi) = (\mu_i(\pi))_{i \in \Pi} \in \RR^\Pi$ of the players' \emph{payoffs}.
	\end{itemize}
\end{definition}

An initialised game is a game in which an initial vertex has been specified.

\begin{definition}[Initialised game]
	An \emph{initialised game} is a tuple $(\Game, v_0)$, often written $\Game_{\|v_0}$, where $\Game$ is a non-initialised game and $v_0 \in V$ is a vertex called \emph{initial vertex}.
\end{definition}

When the context is clear, we often use the word \emph{game} for both initialised and non-initialised games.
Unless stated otherwise, we always make the assumption that the games considered are \emph{finite}, meaning that the underlying graph is finite.

\begin{definition}[Play, history]
	A \emph{play} (resp. \emph{history}) in the game $\Game$ is an infinite (resp. finite) path in the graph $(V, E)$.
	It is also a play (resp. history) in the initialised game $\Game_{\|v_0}$, if $v_0$ is its first vertex.
	The set of plays (resp. histories) in the game $\Game$ (resp. the initialised game $\Game_{\|v_0}$) is denoted by $\Plays \Game$ (resp. $\Plays \Game_{\|v_0}, \Hist \Game, \Hist \Game_{\|v_0}$).
	We write $\Hist_i \Game$ (resp. $\Hist_i \Game_{\|v_0}$) for the set of histories in $\Game$ (resp. $\Game_{\|v_0}$) of the form $hv$, where $v$ is a vertex controlled by player $i$.
\end{definition}

In a game, players select their actions and generate plays by following \emph{strategies}.

\begin{definition}[Strategy, strategy profile]
	A \emph{strategy} for player $i$ in the initialised game $\Game_{\|v_0}$ is a function $\sigma_i: \Hist_i \Game_{\|v_0} \to V$, such that $v\sigma_i(hv)$ is an edge of $(V, E)$ for every $hv$.
	A history $h$ is \emph{compatible} with a strategy $\sigma_i$ if and only if $h_{k+1} = \sigma_i(h_0 \dots h_k)$ for all $k < |h|-1$ such that $h_k \in V_i$. A play $\pi$ is compatible with $\sigma_i$ if all its prefixes are.

	A \emph{strategy profile} for $P \subseteq \Pi$ is a tuple $\bsigma_P = (\sigma_i)_{i \in P}$, where each $\sigma_i$ is a strategy for player $i$ in $\Game_{\|v_0}$.
    A play or a history is \emph{compatible} with $\bsigma_P$ if it is compatible with each $\sigma_i$ for $i \in P$.
    A \emph{complete} strategy profile, usually written $\bsigma$, is a strategy profile for $\Pi$.
\end{definition}

Exactly one play is compatible with the strategy profile $\bsigma$: we call it its \emph{outcome} and write it $\< \bsigma \>$.\footnote{Note that a strategy profile is defined in an initialised game, hence there is no need to specify an initial vertex.}
    
    When $i$ is a player and when the context is clear, we will often write $-i$ for the set $\Pi \setminus \{i\}$.
    When $\btau_P$ and $\btau'_Q$ are two strategy profiles with $P \cap Q = \emptyset$, we write $(\btau_P, \btau'_Q)$ for the strategy profile $\bsigma_{P \cup Q}$ such that $\sigma_i = \tau_i$ for $i \in P$, and $\sigma_i = \tau'_i$ for $i \in Q$.

In a strategy profile $\bsigma_P$, the domains of the strategies $\sigma_i$ are pairwise disjoint. Therefore, we can consider $\bsigma_P$ as one function: for $hv \in \Hist \Game_{\|v_0}$ such that $v \in \bigcup_{i \in P} V_i$, we liberally write $\bsigma_P(hv)$ for $\sigma_i(hv)$ with $i$ such that $v \in V_i$.

\subsection{Equilibria}

Before moving on to SPEs, let us recall the notion of \emph{Nash equilibrium}, and their variants.
Given a strategy $\sigma_i$ in some initialised game $\Game_{\|v_0}$, a \emph{deviation} of $\sigma_i$ is another strategy $\sigma'_i$ for player $i$ in $\Game_{\|v_0}$.
It is a \emph{finite deviation} if the set $\{h \in \Hist_i\Game_{\|v_0} \mid \sigma_i(h) \neq \sigma'_i(h)\}$ is finite.
It is a \emph{one-shot deviation} if that set is empty or a singleton.

\begin{definition}[Nash equilibrium and variants]
	Let $\Game_{\|v_0}$ be a game.
    A strategy profile $\bsigma$ is a \emph{Nash equilibrium} (resp. \emph{weak Nash equilibrium}, resp. \emph{very weak Nash equilibrium})---or \emph{NE} (resp \emph{weak NE}, resp. \emph{very weak NE}) for short---in $\Game_{\|v_0}$ if and only if for each player $i$ and for every deviation (resp. finite deviation, resp. one-shot deviation) $\sigma'_i$ of $\sigma_i$, we have the inequality $\mu_i\< \sigma'_i, \bsigma_{-i} \> \leq \mu_i\< \bsigma \>$.
\end{definition}

To define SPEs, we also need the notions of subgame and substrategy.

\begin{definition}[Subgame, substrategy]
	Let $hv$ be a history in the game $\Game$. The \emph{subgame} of $\Game$ after $hv$ is the initialised game $\Game_{\|hv} = \left(\Pi, V, (V_i)_i, E, \mu_{\|hv}\right)_{\|v}$, where $\mu_{\|hv}$ is the payoff function defined by $\mu_{\|hv}(\pi) = \mu(h\pi)$ for every $\pi \in \Plays \Game_{\|v}$.
    
    If $\sigma_i$ is a strategy in $\Game_{\|v_0}$, its \emph{substrategy} after $hv$ is the strategy $\sigma_{i\|hv}$ in $\Game_{\|hv}$, defined by $\sigma_{i\|hv}(h') = \sigma_i(hh')$ for every $h' \in \Hist_i \Game_{\|hv}$.
\end{definition}

\begin{remark}
    The initialised game $\Game_{\|v_0}$ is also the subgame of $\Game$ after the one-vertex history $v_0$.
\end{remark}

We can now define SPEs.

\begin{definition}[Subgame-perfect equilibrium and variants]
	Let $\Game_{\|v_0}$ be a game.
	The strategy profile $\bsigma$ is a \emph{subgame-perfect equilibrium} (resp. weak subgame-perfect equilibrium, resp. very weak subgame-perfect equilibrium)---or \emph{SPE} (resp. weak SPE, very weak SPE) for short---in $\Game_{\|v_0}$ if and only if for every history $h$ in $\Game_{\|v_0}$, the strategy profile $\bsigma_{\|h}$ is an NE (resp. weak NE, resp. very weak NE) in the subgame $\Game_{\|h}$.
\end{definition}

The following facts are well-known.

\begin{proposition}[\cite{existenceWSPE}]\label{prop:weak_very_weak}
    Every NE is a weak NE and every weak NE is a very weak NE.
    Consequently, every SPE is a weak SPE and every weak SPE is a very weak SPE.
    There is a game with a very weak NE that is not a weak NE, one with a weak NE that is not an NE, and one with a weak SPE that is not an SPE.
    On the other hand, every very weak SPE is a weak SPE.
\end{proposition}

The notion of SPE can be seen as a refinement of Nash equilibrium: it is a stronger equilibrium which excludes players resorting to non-credible threats.

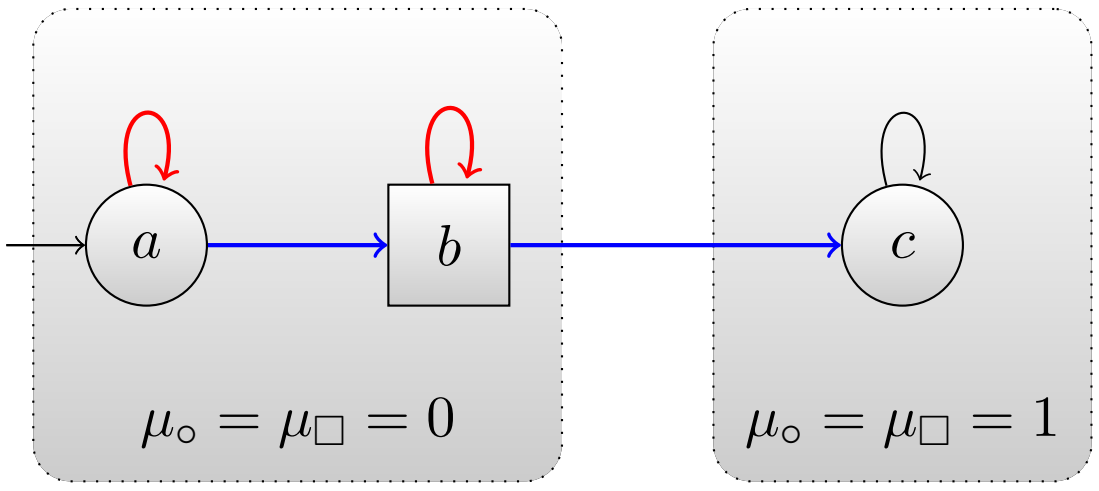
\begin{figure}
    \centering
    \begin{tikzpicture}[auto, draw, node distance=3cm, ->, initial text={}]
        \node[state,rectangle,rounded corners=7pt,dotted,minimum size=2.5cm] (aa) at (1,0) {$\begin{matrix}~\\~\\~\\~\\~~~~~~\mu_\playcircle = \mu_\square = 0~~~~~~\end{matrix}$};
        \node[state, initial] (a) at (0, 0) {$a$};
        \node[state, rectangle] (b) at (2,0) {$b$};
        \node[state,rectangle,rounded corners=7pt,dotted,minimum size=2.5cm, right of=b] (cc) {$\begin{matrix}~\\~\\~\\~\\\mu_\playcircle = \mu_\square = 1\end{matrix}$};
        \node[state] (c) at (5,0) {$c$};
        \path[thick, red] (a) edge[loop above] (a);
        \path[thick, red] (b) edge[loop above] (b);
        \path (c) edge[loop above] (c);
        \path[thick, blue] (a) edge (b);
        \path[thick, blue] (b) edge (c);
    \end{tikzpicture}
    \caption{Two Nash equilibria}
    \label{fig_ne_spe}
\end{figure}

\begin{example} \label{ex_ne_spe}
In the game depicted by Figure~\ref{fig_ne_spe}, where the square vertex is controlled by player $\square$ and the round vertices by player $\Circle$, if both players get the payoff $1$ by reaching the vertex $c$ and $0$ in the other cases, there are actually two memoryless NEs: 
\begin{itemize}
    \item one, depicted in blue, where $\Circle$ goes to the vertex $b$ and then player $\square$ goes to $c$, and both get the payoff $1$;
    \item and one, depicted in red, where player $\Circle$ stays in $a$, and has no incentive to deviate to $b$ because player $\square$ plans to stay in $b$.
\end{itemize}
However, only the first one is an SPE.
Indeed, in the second one, in the subgame after the history $ab$, player $\square$ has a profitable deviation by going to $c$: this is a one-shot deviation, hence that strategy profile is not even a very weak SPE.
\end{example}

\subsection{Mean-payoff games}



		


Mean-payoff games are quantitative games: each action grants each player an immediate reward, and the final payoff of player $i$ is defined as the limit inferior over $n$ (since the limit may not be defined) of the average payoff after $n$ steps.

	\begin{definition}[Mean-payoff, mean-payoff game]
	    In a graph $(V, E)$, we associate to each \emph{reward function} $r: E \to \Q$ the \emph{mean-payoff function}:
	    $$\MP_r: h_0 \dots h_n \mapsto \frac{1}{n} \underset{k=0}{\overset{n-1}{\sum}} r\left(h_k h_{k+1}\right).$$
		A game $\Game = \left(\Pi, V, (V_i)_i, E, \mu \right)$ is a \emph{mean-payoff game} if its underlying graph is finite, and if there exists a tuple $(r_i)_{i \in \Pi}$ of reward functions, such that for each player $i$ and every play $\pi$, we have:
		$$\mu_i(\pi) = \liminf_{n \to \infty} \MP_{r_i}(\pi_{\leq n}).$$
	\end{definition}
	
When the context is clear, we liberally write $\MP_i(h)$ for the quantity $\MP_{r_i}(h)$, and $\MP(h)$ for the tuple $(\MP_i(h))_i$, as well as $r(uv)$ for $(r_i(uv))_i$.

\begin{example}
    The game of Figure~\ref{fig_ne_spe} 
    can be seen as a mean-payoff game: if the rewards $r_\playcircle(aa)$, $r_\square(aa)$, $r_\playcircle(bb)$, and $r_\square(bb)$ are defined as equal to $0$, and the rewards $r_\playcircle(cc)$ and $r_\square(cc)$ as equal to $1$, no matter what are the other ones, both players will get the payoff $1$ if they reach the vertex $c$, and the payoff $0$ otherwise.
\end{example}

\paragraph{Prefix-independent and prefix-linear games.}

A useful property of 
mean-payoff games is that they are 
\emph{prefix-independent}.

\begin{definition}[Prefix-independent game]
	A game $\Game$ is \emph{prefix-independent} if, for every history $h$ and for every play $\pi$, we have $\mu(h\pi) = \mu(\pi)$.
	We also say, in that case, that the payoff function $\mu$ is prefix-independent.
\end{definition}




In this work, we will however consider a broader class: \emph{prefix-linear} games, as defined in~\cite{DBLP:journals/iandc/RouxP18}.

\begin{definition}[Prefix-linear game]
	A game $\Game$ is \emph{prefix-linear} if, for every history $h$ and for every plays $\pi, \pi'$, for each player $i$, we have $\mu_i(\pi) < \mu_i(\pi')$ if and only if $\mu_i(h \pi) < \mu_i(h \pi')$.
	We also say, in that case, that the payoff function $\mu$ is prefix-linear.
\end{definition}

\begin{remark}
    Every prefix-independent game is prefix-linear.
\end{remark}

\subsection{The constrained existence problem}

For every equilibrium notion $\Eq$ (NE, SPE, and their variants), we study the complexity of the following decision problem.

\begin{definition}[$\Eq$ constrained existence problem]
    Given a finite game $\Game_{\|v_0}$ and two payoff profiles $\bx, \by \in \Q^\Pi$, does there exist an $\Eq$-equilibrium $\bsigma$ in $\Game_{\|v_0}$ such that $\bx \leq \mu\< \bsigma \> \leq \by$?
\end{definition}

\section{The Negotiation Function}\label{sec:nego}

In this section, we present the negotiation function: its definition, its fundamental properties, and its history, with the main results it has already led to.

		\subsection{Requirements}

The negotiation function takes as input a \emph{requirement}, that is, a vertex labelling that indicates, on which vertex, the minimal payoff the player controlling that vertex can accept while punishing another player.

\begin{definition}[Requirement]
	A \emph{requirement} on the game $\Game$ is a mapping $\lambda: V \to \RR \cup \{\pm\infty\}$.
\end{definition}

For a given vertex $v$, the quantity $\lambda(v)$ represents the minimal payoff that the player controlling $v$ will require in a play starting from $v$.

\begin{definition}[$\lambda$-consistency]\label{def:lambda_cons}
	Let $\lambda$ be a requirement on a game $\Game$.
 A play $\pi$ in the game $\Game$ is \emph{$\lambda$-consistent} if and only if, for every player $i \in \Pi$ and integer $n \in \NN$ with $\pi_n \in V_i$, we have $\mu_i(\pi_{\geq n})~\geq~\lambda(\pi_n)$.
	The set of $\lambda$-consistent plays from a vertex $v$ is denoted by $\lCons\Game_{\|v}$, or $\lCons(v)$ when the context is clear.
\end{definition}

Such a requirement also induces a notion of rationality for strategy profiles aiming at punishing one player.

\begin{definition}[$\lambda$-rationality] \label{def:lambda_rat}
	Let $\lambda$ be a requirement on a game $\Game$.
 Let $i \in \Pi$.
 The strategy profile $\bsigma_{-i}$ is \emph{$\lambda$-rational} if and only if for every history $hv$ compatible with $\bsigma_{-i}$, there is a $\lambda$-consistent play that is compatible with $\bsigma_{-i\|hv}$.
	We denote by $\lRat_{-i}\Game_{\|v}$, or $\lRat(v)$ when the context is clear, the set of $\lambda$-rational strategy profiles for $-i$ in the game $\Game_{\|v}$.\footnote{Other papers about the negotiation function use a different definition, with the existence of a strategy for player $i$ that generates $\lambda$-consistent plays.
    These two definitions are equivalent.}
\end{definition}
	
Note that $\lambda$-rationality is a property of a strategy profile for all the players but one, player $i$. Intuitively, their rationality is justified by the fact that they collectively assume that player $i$ will, eventually, play according to the strategy $\sigma_i$: if it is the case, then everyone gets their payoff satisfied.

Finally, let us define a particular requirement: the \emph{vacuous requirement}, that requires nothing, and with which every play is consistent.

\begin{definition}[Vacuous requirement]\label{def:lambda0}
	In any game, the \emph{vacuous requirement}, denoted by $\lambda_0$, is the requirement constantly equal to $-\infty$.
\end{definition}

		\subsection{Negotiation} \label{ss_def_nego}

The negotiation function captures a negotiation process: when a player has a requirement to satisfy, another player can hope for a better payoff than what they can secure in general, and therefore update their own requirement.
Note that we always use the convention $\inf \emptyset = +\infty$.

\begin{definition}[Negotiation function, steady negotiation] \label{def_nego}
	Let $\Game$ be a game.
	The \emph{negotiation function} is the function that transforms each requirement $\lambda$ on $\Game$ into a requirement $\nego(\lambda)$ on $\Game$, such that for each player $i \in \Pi$ and every vertex $v \in V_i$, we have:
	$$\nego(\lambda)(v) = \inf_{\bsigma_{-i} \in \lRat_{-i}(v)} \sup_{\sigma_i} \mu_i\< \bsigma\>.$$
	If that infimum is realised for every requirement $\lambda$, player $i$ and vertex $v \in V_i$ such that $\lRat(v) \neq \emptyset$, then the game $\Game$ is called a game \emph{with steady negotiation}.
\end{definition}

The quantity $\nego(\lambda)(v)$ is then the worst case value that the player controlling $v$ can ensure, assuming that the other players play $\lambda$-rationally.

\begin{remark}\label{rk:nego_properties}
    The negotiation function satisfies the following properties.
    
    \begin{itemize}
        \item It is monotone: if we have $\lambda \leq \lambda'$, then we have $\nego(\lambda) \leq \nego(\lambda')$.
        
        \item It is also inflationary: for every $\lambda$, we have $\lambda \leq \nego(\lambda)$.
        
        \item There exists a $\lambda$-rational strategy profile from $v$ against the player controlling $v$ if and only if $\nego(\lambda)(v) \neq +\infty$.
    \end{itemize}
\end{remark}

\begin{example}    
    Let us consider again the game depicted by Figure~\ref{fig_ne_spe}, and let us consider the vacuous requirement $\lambda_0: v \to -\infty$.
    Let us compute $\lambda_1 = \nego(\lambda_0)$.
    
    From the vertex $c$, the only payoff that can be obtained is $\lambda_1(c) = 1$.
    From the vertex $b$, player $\square$ can go to the vertex $c$ and get the payoff $\lambda_1(b) = 1$.
    From the vertex $a$, even if player $\Circle$ goes to $b$, player $\square$ can stay there, hence she cannot get more than $\lambda_1(a) = 0$.

    Let us now compute $\lambda_2 = \nego(\lambda_1)$.
    From $c$, the only payoff possible is still $\lambda_2(c) = 1$, and from $b$, it is still possible to get $\lambda_2(b) = 1$ and nothing more.
    But from the vertex $a$, a $\lambda$-rational strategy profile must necessarily plan to go to the vertex $c$ if the vertex $b$ is reached, hence $\lambda_2(a) = 1$.
    
    Finally, we have $\nego(\lambda_2) = \lambda_2$.
\end{example}

        \subsection{Negotiation and equilibria}

The negotiation function provides the correct grammar to rephrase a folklore characterisation of Nash equilibria outcomes: a play is a Nash equilibrium outcome if and only if it gives to every player at least the payoff that they could enforce by deviating, if all the other players try to punish them---i.e., if for every vertex it traverses, it gives to the player controlling that vertex its adversarial value.

\begin{theorem}[\cite{LMCS23}]\label{prop_ne}
	Let $\Game$ be a prefix-independent game with steady negotiation, and let $\pi$ be a play in the game $\Game$.
    Then, the play $\pi$ is an NE outcome if and only if $\pi$ is $\nego(\lambda_0)$-consistent.
\end{theorem}


However, the real power of the negotiation function lies in the fact that its fixed points, in a similar way, characterise subgame-perfect equilibria.

\begin{theorem}[\cite{LMCS23}]\label{thm:nego_spe}
	Let $\Game$ be a prefix-independent game with steady negotiation, and let $\pi$ be a play in the game $\Game$.
    Then, the play $\pi$ is an SPE outcome if and only if $\pi$ is $\lambda$-consistent, for some fixed point $\lambda$ of the negotiation function.
\end{theorem}


\begin{example}
    Let us consider again the game of Figure~\ref{fig_ne_spe}.
    The outcomes of the two NEs that we identified in Example~\ref{ex_ne_spe} are indeed $\lambda_1$-consistent: the play $a^\omega$ and the play $abc^\omega$.
    However, only the latter is $\lambda_2$-consistent, and $\lambda_2$ is a fixed point of the negotiation function.
\end{example}

\subsection{History of the concept and applications}

\paragraph{Antecedents.}
Prior to its formalisation, the idea behind the concept of negotiation seems to have emerged in two parallel ways.

On the one hand, János Flesch and Arkadi Predtetchinski introduced in 2017 a characterisation of SPEs, based on an iterative process where labels on histories are updated through computing the value of an auxiliary game~\cite{DBLP:journals/mor/FleschP17}.
One iteration of that process corresponds to what was later called the negotiation function, and the auxiliary game to the \emph{abstract negotiation game}~\cite{LMCS23}.
This characterisation applies to Borel games where the payoff functions have finite range.
When they have an infinite range (like in mean-payoff games), their work only proves that the process characterises a set of plays that is known to contain only outcomes of $\epsilon$-SPEs, for every $\epsilon > 0$, as well as all outcomes of SPEs.
It was later shown~\cite{LMCS23} that it actually characterises SPE outcomes in all games with steady negotiation.
Since their work was primarily conducted with the objective of proving existence results, Flesch and Predtetchinski's approach does not provide a characterisation that can be directly used for algorithmic purposes, as their iterative process may be infinite, and each iteration requires solving an infinite number of two-player zero-sum games, which can themselves be of infinite size.

On the other hand, in 2019, Marie van den Bogaard, Thomas Brihaye, Véronique Bruyère, Aline Goeminne, and Jean-François Raskin studied SPEs in \emph{quantitative reachability games}, where players aim to reach a given set of vertices as quickly as possible~\cite{DBLP:journals/lmcs/BrihayeBGRB20}---note that those games do not have finite range, and are therefore not covered by Flesch et al.'s full characterisation.
Their focus was computational: they showed $\PSpace$-completeness of the constrained existence problem of SPEs in those games, that is, the problem of deciding, given such a game and a collection of intervals for each player's payoff, whether the game contains an SPE where each player's payoff lies in the specified interval.
Their upper bound relies on an algorithm that labels each vertex with an integer, and updates this labelling iteratively; these updates, which correspond to one iteration of the negotiation function, continue until reaching a fixed point, which characterises SPEs.

In both cases, the negotiation function was not explicitly formalised as a function but was instead an implicit part of an iterative process.
However, reasoning about these iterations as a function, whose fixed points characterise SPEs, proved to be useful in cases where simply iterating the function is not the most efficient way to solve our problem, or could even fail to terminate, as is the case in mean-payoff games.

\paragraph{The negotiation function.}

The negotiation function was defined first for the study of mean-payoff games (in~\cite{Concur}, and its journal version~\cite{LMCS23}).
In this work, it was shown that, again, SPEs outcomes can be characterised as the plays that are consistent with the least fixed point of some iterative process; but contrary to quantitative reachability games, this iterative process is not guaranteed to terminate in mean-payoff games.
In order to show decidability for the constrained existence problem in that setting, the negotiation function needed to be studied as such.
This function is piecewise affine, and one can construct an explicit representation of it, which makes it possible to extract its least fixed point: this algorithm places the constrained existence problem in $\ExpTime$.
The same paper shows that the negotiation function characterises SPEs in all prefix-independent games with steady negotiation.
Moreover, the negotiation function can also characterise a quantitative relaxation of SPEs, known as $\epsilon$-SPEs, whose constrained existence problem is also decidable.

The negotiation function was used later in parity games~\cite{CSL}.
There, the constrained existence problem of SPEs was known to be decidable, but a complexity gap remained between an $\NP$ lower bound and an $\ExpTime$ upper bound~\cite{GU08}.
Iterating the negotiation function until a fixed point is reached does terminate in those games, but also yields an exponential-time algorithm.
However, SPEs are not only characterised by the least fixed point of the negotiation function: a play is an SPE outcome if and only if it is consistent with \emph{some} of its fixed points.
It is thus sufficient to guess a vertex labelling, to check whether it is a fixed point of the negotiation function, and that some play compatible with it matches the desired constraints.
The constrained existence problem was thus shown to be $\NP$-complete.
Later on, this idea was used on mean-payoff games, also closing the complexity gap that was left earlier and showing $\NP$-completeness~\cite{Icalp}.
These results were recently gathered in one journal paper~\cite{jacm}.

One last application of the negotiation function, again in mean-payoff games, was for a different but close problem, the \emph{achaotic rational verification problem}~\cite{mfcs23}, which we can phrase as follows: given a game, a player, and a threshold, is it the case that the player always gets a payoff above the threshold in every $\epsilon$-SPE, where $\epsilon$ is such that $\epsilon$-SPE do exist?
Since SPEs can be understood as a notion of collective rationality, but are not guaranteed to exist in mean-payoff games, this problem intuitively asks whether the specified player is guaranteed a sufficient payoff in all strategy profiles that are \emph{as rational as they can be}.
This problem is $\PTime^\NP$-complete, and the upper bound uses, again, the negotiation function.
Interestingly, the analysis of the negotiation function was also used to show that there exists a minimal $\epsilon$ such that $\epsilon$-SPEs exist, and that this value has polynomially bounded bit-size: although this was proved as an intermediate result toward the complexity upper bound, it can be considered a first not immediately computational application of the negotiation function.

All those results can be found in the author's PhD thesis~\cite{mythesis}.




\section{The weak negotiation function}\label{sec:weak_nego}

In this section, we show that a variant of the negotiation function can handle weak SPEs in the same way the negotiation function handles SPEs.
That variant is much simpler to define than the standard negotiation function, since it does not require an equivalent of the notion of $\lambda$-rationality: since weak SPEs coincide with very weak SPEs, we only need to consider one-shot deviations, and the weak negotiation function can be defined as the worst payoff that can be imposed on a player after such a deviation, in a $\lambda$-consistent play.

\subsection{Definition and first properties}

\begin{definition}[Weak negotiation function]
    For each requirement $\lambda$, we define the requirement $\wnego(\lambda)$ that maps each vertex $v$ to the value:
    $$\wnego(\lambda)(v) = \max_{vw \in E} \inf_{\pi \in \lCons(w)} \mu_i(v\pi),$$
    where $i$ is the player controlling $v$.
\end{definition}

Let us first note some straightforward properties of this function, analogous to \Cref{rk:nego_properties}.

\begin{remark}\label{rk:wnego_properties}
    The weak negotiation function satisfies the following properties.
    
    \begin{itemize}
        \item It is monotone: if we have $\lambda \leq \lambda'$, then we have $\wnego(\lambda) \leq \wnego(\lambda')$.
        Therefore, the set of requirements is a complete lattice, and by Tarski's fixed point theorem, it has a least fixed point $\lambda^\star$.

        \item It is an under-approximation of the negotiation function: for every $\lambda$, we have $\wnego(\lambda) \leq \nego(\lambda)$.
    \end{itemize}
\end{remark}

On the other hand, the weak negotiation function is not necessarily inflationary, but we have the following property.

\begin{lemma}\label{lm:if_non_increasing}
    If we have $\wnego(\lambda) \leq \lambda$, then we have $\lambda \geq \lambda^\star$.
\end{lemma}

\begin{proof}
    Let $\Lambda$ be the set of requirements satisfying $\wnego(\lambda) \leq \lambda$.
    That set is nonempty; let $\lambda_\bot$ be its infimum.
    The set $\Lambda$ contains all fixed points of the weak negotiation function, and in particular $\lambda^\star$; therefore, we have $\lambda_\bot \leq \lambda^\star$.
    
    On the other hand, for every $\lambda \in \Lambda$ we have $\lambda_\bot \leq \lambda$, and therefore $\wnego(\lambda_\bot) \leq \wnego(\lambda) \leq \lambda$.
    By taking the infimum over $\lambda$ we get $\wnego(\lambda_\bot) \leq \lambda_\bot$, i.e. $\lambda_\bot \in \Lambda$.
    But $\wnego(\lambda_\bot) \leq \lambda_\bot$ entails $\wnego(\wnego(\lambda_\bot)) \leq \wnego(\lambda_\bot)$, i.e., $\wnego(\lambda_\bot) \in \Lambda$, and since $\lambda_\bot$ is the smallest element of $\Lambda$, this gives $\lambda_\bot \leq \wnego(\lambda_\bot)$, hence $\lambda_\bot$ is a fixed point of $\wnego$.
    Therefore $\lambda_\bot \geq \lambda^\star$.

    We obtain $\lambda_\bot = \lambda^\star$, and therefore $\lambda \geq \lambda^\star$ for every $\lambda \in \Lambda$.
\end{proof}

\subsection{Steady weak negotiation}

We can also define an equivalent of the notion of steady negotiation.

\begin{definition}[Games with steady weak negotiation]
    A game is with \emph{steady weak negotiation} if and only if for every requirement $\lambda$, every player $i$ and every edge $vw$ with $v \in V_i$, if there exists a $\lambda$-consistent play $\pi$ from $w$, there exists one with $\mu_i(v\pi) \leq \wnego(\lambda)(v)$.
\end{definition}

Technically, a game with steady negotiation may not be with steady weak negotiation.
It is however the case if we slightly strengthen the steady negotiation assumption.

\begin{definition}
    Let $\Game$ be a game.
    A \emph{spied version} of $\Game$ is a game $\Game'$ where, for some player $i$ and some edge $vw \in E$ with $v \in V_i$, one has added:
    \begin{itemize}
        \item a fresh player $i'$;

        \item a fresh vertex $v'$ controlled by $i'$;
        
        \item an edge $v'w$;
    \end{itemize}
    with $\mu(v'\pi) = \mu(v\pi)$ for every play $v'\pi$, and $\mu_{i'} = \mu_i$.
\end{definition}

\begin{remark}
    A spied version of a parity game is a parity game.
    A spied version of a mean-payoff game is a mean-payoff game.
\end{remark}

We can then make the following observation.

\begin{lemma}\label{lm:steady_steadyw}
    Let $\Game$ be a game such that every spied version of $\Game$ has steady negotiation.
    Then, the game $\Game$ has steady weak negotiation.
\end{lemma}

\begin{proof}
    Let $\lambda$ be a requirement in $\Game$.
    Let $vw$ be an edge, with $v \in V_i$, and let us assume that $\lCons(w)$ is nonempty.
    We want to show that there exists a $\lambda$-consistent play $\pi$ from $w$ such that $\mu_i(v\pi) \leq \wnego(\lambda)(v)$.

    Let $\Game'$ be a spied version of $\Game$ obtained by duplicating the edge $vw$.
    Since $\Game'$ is with steady negotiation, by setting $\lambda(v') = -\infty$, there exists a $\lambda$-rational strategy profile $\bsigma_{-i'}$ from $v'$ that minimises the quantity $\sup_{\sigma_{i'}} \mu_{i'}\< \bsigma \>$.
    Since player $i'$ never has any non-trivial choice to make in $\Game'$, the play $\< \bsigma \> = v'\pi$ does not depend on $\sigma_{i'}$, hence it is $\lambda$-consistent, and it minimises player $i'$'s payoff over all $\lambda$-consistent plays starting with the edge $v'w$.
    Therefore, the play $v\pi$ minimises player $i$'s payoff over all $\lambda$-consistent plays starting with the edge $vw$, hence $\mu_i(v\pi) \leq \wnego(\lambda)(v)$, as desired.
\end{proof}

This lemma shows that, for all practical purposes, steady negotiation is stronger than steady weak negotiation.
In particular, it shows that mean-payoff games, which are known to have steady negotiation~\cite{jacm}, have steady weak negotiation.

\subsection{A characterisation of weak SPEs}

We are now equipped to prove the following theorem, analogous to \Cref{thm:nego_spe}.
Moreover, we prove it in a (for all practical purposes) broader class, since instead of prefix-independence, we only need prefix-linearity, and instead of steady negotiation, we only need steady weak negotiation.

\begin{theorem} \label{thm:wnego}
    Let $\Game$ be a prefix-linear game, and let $\pi$ be a play in $\Game$.
    If $\pi$ in $\Game$ is a weak SPE outcome, then it is $\lambda$-consistent for some fixed point $\lambda$ of the weak negotiation function.
    The converse is true if $\Game$ has steady weak negotiation.
\end{theorem}

\begin{proof}
    We first prove that all weak SPE outcomes are consistent with some fixed point of the weak negotiation function, and then the converse.

    \paragraph{If $\pi$ is a weak SPE outcome, then there is a fixed point $\lambda$ of $\wnego$ such that $\pi$ is $\lambda$-consistent.}

        Let us consider a weak SPE $\bsigma$ in the game $\Game_{\|v_0}$, such that $\< \bsigma \> = \pi$.
        Let us define the requirement:
        $$\lambda_{\bsigma}: v \mapsto \inf_{hv \in \Hist \Game_{\|v_0}} \mu_i\< \bsigma_{\|hv} \>,$$
        where $i$ is the player controlling the vertex $v$.
        By construction, every play of the form $\< \bsigma_{\|hv}\>$, and in particular $\pi$ itself, is $\lambda_{\bsigma}$-consistent.
        We now show the following.

        \begin{claim}
            We have $\wnego(\lambda_{\bsigma}) \leq \lambda_{\bsigma}$.
        \end{claim}

        \begin{proof}
            Let $i \in \Pi$ and $v \in V_i$.
            Let us assume that $\wnego(\lambda_{\bsigma})(v) > \lambda_{\bsigma}(v)$: then, there exists an edge $vw \in E$ such that we have:
            $$\inf_{\chi \in \lambda_{\bsigma}\Cons(w)} \mu_i(v\chi) > \lambda_{\bsigma}(v) = \inf_{hv \in \Hist \Game_{\|v_0}} \mu_i\< \bsigma_{\|hv} \>,$$
            and in particular:
            $$\inf_{hv \in \Hist \Game_{\|v_0}} \mu_i(v \< \bsigma_{\|hvw} \>) > \inf_{hv \in \Hist \Game_{\|v_0}} \mu_i\< \bsigma_{\|hv} \>,$$
            which means that there exists a history $hv$ such that $\mu_i(v\< \bsigma_{\|hvw} \>) > \mu_i\< \bsigma_{\|hv} \>$.
            By prefix-linearity, this implies $\mu_i(h v \< \bsigma_{\|hvw} \>) > \mu_i(h \< \bsigma_{\|hv} \>)$, which is impossible since $\bsigma$ is a weak SPE.
        \end{proof}
        
        This is not enough to conclude that $\lambda_{\bsigma}$ is a fixed point of the weak negotiation function, since $\wnego$ is not necessarily inflationary; but by \Cref{lm:if_non_increasing}, it gives $\lambda_{\bsigma} \geq \lambda^\star$, and therefore $\pi$ is $\lambda^\star$-consistent.

\paragraph{If $\pi$ is $\lambda$-consistent for some fixed point $\lambda$ of $\wnego$ and if $\Game$ has steady weak negotiation, then $\pi$ is a weak SPE outcome.}

            Let us define the strategy profile $\bsigma$ inductively, maintaining the induction hypothesis that all plays of the form $\< \bsigma_{\|h} \>$, for all histories $h$, are $\lambda$-consistent.
            First, we define $\< \bsigma \> = \pi$, which satisfies the induction hypothesis.
            
            Then, let $hvw$ be a history such that $\< \bsigma_{\|hv}\>$ is defined but $\< \bsigma_{\|hvw} \>$ is not.
            Let $i$ be the player controlling the vertex $v$.
            By the induction hypothesis, the play $\< \bsigma_{\|hv} \>$ is $\lambda$-consistent.
            Therefore, we have $\mu_i\< \bsigma_{\|hv} \> \geq \lambda(v)$.
            Since $\lambda$ is a fixed point of the weak negotiation function, we have
            $\mu_i\< \bsigma_{\|hv} \> \geq \wnego(\lambda)(v)$;
            since $\wnego(\lambda) \neq +\infty$ we have $\lCons(w) \neq \emptyset$, and since $\Game$ is a game with steady weak negotiation, there exists a play $\chi \in \lCons(w)$ with $\mu_i(v\chi) \leq \wnego(\lambda)(v)$.
            We define $\< \bsigma_{\|hvw} \> = \chi$, which satisfies the induction hypothesis.

            By construction, the strategy profile $\bsigma$ is a very weak SPE.
            By \Cref{prop:weak_very_weak}, it is therefore a weak SPE.
\end{proof}

In the rest of this document, we will use this theorem in mean-payoff games.
Let us however mention that as a characterisation, it also applies to a wide range of known game classes such as parity games, total payoff games, or discounted-sum games.

\section{Algorithms} \label{sec:algos}

From \Cref{thm:wnego}, we can obtain straightforward complexity results for the constrained existence problem of weak SPEs.
In parity games, the constrained existence problem of weak SPEs is already known to be $\NP$-complete~\cite{DBLP:journals/iandc/BrihayeBGR21}.
We prove that this result also holds in mean-payoff games, thus finding the same complexities as for SPEs~\cite{jacm}.

\subsection{Sizes}

For each integer $n$, we write $\lv n \rv = \lceil \log_2(|n|+1) \rceil + 2$ for the \emph{bit-size} of $n$, and for each rational $p/q$ where $p$ and $q$ are co-prime, we write $\lv p/q \rv = \lv p \rv + \lv q \rv + 1$.
Where $z$ is irrational, we set $\lv z \rv = +\infty$, and when $z = \pm \infty$, we set $\lv z \rv = 1$.
The bit-size of a set, or of a tuple, is defined as the sum of the bit-sizes of its elements, plus $1$; in particular, every mean-payoff game is assumed represented by a table of rewards, and therefore has finite bit-size.

\subsection{The size of the least fixed point}

Throughout the remainder of this section, we assume given a mean-payoff game $\Game_{\|v_0}$.
We denote the least fixed point of the weak negotiation function on that game by $\lambda^\star$.
The main technical difficulty is in proving the following lemma: the requirement $\lambda^\star$ can be described with polynomially many bits.

\begin{lemma}\label{lm:size_fixed_point}
    There exists a polynomial $P$, independent of $\Game, \bx, \by$, such that we have $\lv \lambda^\star \rv \leq P(\lv \Game \rv)$.
\end{lemma}

\begin{proof}
    The following arguments are inspired by the proof of the analogous result about SPEs~\cite[Lemma~5.10]{jacm}.

    \paragraph{Feasible plays.}
    For each set $I \subseteq V$ of vertices, consider the set $F_{I}$ of payoff vectors $\mu(\pi)$ of plays $\pi$ in $\Game$ that satisfy $\Inf(\pi) = I$.
    It is known~\cite{DBLP:conf/concur/ChatterjeeDEHR10} that this set has the form:
    $$F_I = \dseal \left(\underset{c \in \SC(I, E \cap (I \times I))}{\Conv} \mu(c^\omega)\right),$$
    where $\SC(I, E \cap (I \times I))$ denotes the set of simple cycles in the graph $(I, E \cap (I \times I))$, and $\dseal X$, for every set $X \subseteq \RR^\Pi$, is the \emph{downward sealing} of $X$, i.e. the set:
    $$\dseal X = \left\{\left. \left( \min_{\bz \in Z} z_i\right)_{i \in \Pi} ~\right|~ Z \text{ is a finite subset of } X\right\}.$$
    It has been shown~\cite[Proof of Lemma~5.10]{jacm} that there is a polynomial $P_1$, independent of $\Game$, such that every such set is a polytope that can be described by a system of inequalities whose size are bounded by $P_1(\lv \Game \rv)$.
    (Note that this does not say anything about the number of inequalities in that system.)

    \paragraph{Consistent plays.}
    Now, for a given requirement $\lambda$ and a set $O$ of vertices, we can define the set $C_{IO}^\lambda$ of payoff vectors of $\lambda$-consistent plays $\pi$ such that $\Inf(\pi) = I$ and $\Occ(\pi) = O$: if there is at least such a play, then the set $C_{IO}$ is a polytope defined by the same inequalities that define $F_I$, plus the inequality $x_i \geq \lambda(v)$ for each player $i$ and vertex $v \in O \cap V_i$.
    Otherwise, it is the empty set, defined by a trivial inequality.

    \paragraph{Product construction.}
    Consider now, in the space $\RR^{V \times \Pi}$, the set:
    $$X_\lambda = \prod_{v \in V} \bigcap_{vw \in E} \bigcup_{\substack{O, I \\ w \in O}} \uparrow C_{IO}^\lambda,$$
    where $\uparrow Y$ is the upward closure of the set $Y$.
    It is a union of polytopes that are each defined by inequalities of the forms mentioned above.

    For each point $\bbx \in \RR^{V \times \Pi}$, define the requirement $\lambda_{\bbx}$ by $\lambda_{\bbx}(v) = x_{vi}$ for each $i$ and each $v \in V_i$.

    \begin{proposition}\label{prop:wnego_polytopes}
        For each player $i$ each vertex $v \in V_i$, if $\lCons(w) \neq \emptyset$ for every $uw \in E$, then we have:
        $$\wnego(\lambda)(v) = \inf_{\bbx \in X_\lambda} \lambda_{\bbx}(v).$$
    \end{proposition}

    \begin{proof}
        By definition of the weak negotiation function, we have:
        $$\wnego(\lambda)(v) = \max_{vw \in E} \inf_{\pi \in \lCons(w)} \mu_i(v\pi)$$
        $$= \max_{vw \in E} \inf_{\substack{I, O \\ w \in O}} \inf\left\{ \mu_i(\pi)  \mid \pi \in \lCons(w), \Occ(\pi) = O, \Inf(\pi) = I\right\}$$
        $$= \max_{vw \in E} \inf_{\substack{I, O \\ w \in O}} \inf \left\{x_i ~\left|~ \bx \in C_{IO}^\lambda\right.\right\}$$
        $$= \max_{vw \in E} \inf_{\substack{I, O \\ w \in O}} \inf \left\{x_i ~\left|~ \bx \in ~ \uparrow C_{IO}^\lambda\right.\right\}$$
        \begin{equation}
            = \max_{vw \in E}  \inf \left\{x_i ~\left|~ \bx \in \bigcup_{\substack{I, O \\ w \in O}} \uparrow C_{IO}^\lambda\right.\right\}
            \label{eq:beforelast}
        \end{equation}
        \begin{equation}
            =   \inf \left\{x_i ~\left|~ \bx \in \bigcap_{vw \in E} \bigcup_{\substack{I, O \\ w \in O}} \uparrow C_{IO}^\lambda\right.\right\}
            \label{eq:last}
        \end{equation}

        This last line requires a justification.
        Let $v w^\star$ realise the maximum above.
        Since we have:
        $$\bigcap_{vw \in E} \bigcup_{\substack{I, O \\ w \in O}} \uparrow C_{IO}^\lambda \subseteq \bigcup_{\substack{I, O \\ w^\star \in O}} \uparrow C_{IO}^\lambda,$$
        the fact that the line before the quantity \ref{eq:beforelast} is smaller than or equal to the quantity \ref{eq:last} is straightforward.
        
        On the other hand, for every $\bx^{vw^\star} \in \bigcup_{\substack{I, O, w^\star \in O}} \uparrow C_{IO}^\lambda$, since $vw^\star$ realise the maximum above, there exists for each edge $vw \neq vw^\star$ a point $\bx^{vw}$ with $x^{vw}_i \leq x^{vw^\star}_i$.
        If we now take the point $\bx$ defined by $x_j = \max_{vw \in E} x^{vw}_j$ for each $j$, then this point is in the intersection:
        $$\bigcap_{vw \in E} \bigcup_{\substack{I, O \\ w \in O}} \uparrow C_{IO}^\lambda$$
        by upward-closedness of the intersected sets, and it satisfies $x_i = x^{vw^\star}_i$.
        Hence the quantity \ref{eq:last} is smaller than or equal to, and actually equal to, the quantity \ref{eq:beforelast}.

        In the higher-dimensional space $X_\lambda$, which is nonempty since $\lCons(w) \neq \emptyset$ for each edge $vw$ this translates to:
        $$\wnego(\lambda)(v) = \inf_{\bbx \in X_\lambda} x_{vi} = \inf_{\bbx \in X_\lambda} \lambda_{\bbx}(v)$$
    \end{proof}

    \paragraph{Diagonal construction.}
    Let us now define the following set:
    $$\Delta = \left\{\left. \bbx \in \RR^{V \times \Pi} ~\right|~ \bbx \in X_{\lambda_{\bbx}}\right\}.$$
    This set is also a union of polyhedra, defined by inequalities of the same form as the inequalities defining each set $C_{IO}^\lambda$, but where each inequality of the form $x_{ui} \geq \lambda(v)$ is replaced by a conjunction of inequalities of the form $x_{ui} \geq x_{vi}$, which have bounded size, and that we will therefore neglect.
    Therefore, all the inequalities defining those polyhedra have size at most $P(\lv \Game \rv)$.
    Moreover, the set $\Delta$ has the following property.

    \begin{proposition}
        A requirement $\lambda$ with finite values satisfies $\wnego(\lambda) \leq \lambda$ if and only if $\lambda = \lambda_{\bbx}$ for some $\bbx \in \Delta$.
    \end{proposition}

    \begin{proof}
        If $\lambda = \lambda_{\bbx}$ for some $\bbx \in \Delta$, then by definition of $\Delta$ we have $\bbx \in X_\lambda$, and by \Cref{prop:wnego_polytopes}, we have, for each $v$:
        $$\wnego(\lambda)(v) = \inf_{\bby \in X_\lambda} \lambda_{\bby}(v) \leq \lambda_{\bbx}(v) = \lambda(v).$$

        Conversely, if $\wnego(\lambda) \leq \lambda$, then by \Cref{prop:wnego_polytopes} again, we have for each $v$:
        $$\lambda(v) \geq \inf_{\bbx \in X_\lambda} \lambda_{\bbx}(v).$$
        Let us observe that the set $X_\lambda$ is a finite union of closed polyhedra that are bounded below.
        Therefore, the infimum above is realised, and there exists a point $\bby^v \in X_\lambda$ such that $\lambda(v) \geq \lambda_{\bby^v}(v)$.
        Moreover, the set $X_\lambda$ is upward-closed, hence if we define $x^v_{vi} = \lambda(v)$ and $x^v_{wj} = y^v_{wj}$ for every other pair $(w, j)$, then the point $\bbx^v$ also belongs to $X_\lambda$.
        
        Consider now the point $\bbx$ defined by $x_{vi} = x^v_{vi}$ for each $v$ and each $i$.
        Then, we have $\lambda_{\bbx} = \lambda$, and since $X_\lambda$ is a Cartesian product, we have $\bbx \in X_\lambda$.
    \end{proof}

    Thus, the requirement $\lambda^\star$, which is guaranteed to have finite values since mean-payoff games always contain weak SPEs~\cite{existenceWSPE}, is the (unique) minimal element of the set $\{\lambda_{\bbx} \mid \bbx \in \Delta\}$.
    We can therefore write $\lambda^\star = \lambda_{\bbx^\star}$ where $\bbx^\star$ is a point of $\Delta$ that minimises the linear mapping $\bbx \mapsto \sum_{i, u \in V_i} x_{ui}$; and such a point can be chosen as a vertex of one of the polytopes composing $\Delta$.
    By standard results on polytopes defined by inequalities~\cite{DBLP:conf/cav/BrenguierR15,Schrijver1986}, the bit-size of every vertex of a polytope is bounded by a polynomial of the maximal bit-size of the inequalities defining that polytope and of the dimension of the space, hence the result.
\end{proof}

\subsection{Conclusion}

We can now prove our final result.

\begin{theorem} \label{thm:weakSPEs_mp}
    The constrained existence problem of weak SPEs in mean-payoff games is $\NP$-complete.
\end{theorem}

\begin{proof}
    For hardness, we refer to the $\NP$-hardness proof for the constrained existence problem of SPEs in mean-payoff games~\cite[Theorem 8.1]{LMCS23}.
    That proof goes by reduction to a game in which one can observe that all deviations considered are one-shot; it therefore also covers our problem.

    As for easiness, let $\Game$, $\bx$, $\by$ form an instance of the constrained existence problem of weak SPEs in mean-payoff games.
    By \Cref{lm:size_fixed_point,thm:wnego}, it is a positive instance if and only if there exists a requirement $\lambda$ with bit-size at most $P(\lv \Game \rv)$ and a play $\pi$ such that:
    \begin{itemize}
        \item the play $\pi$ is $\lambda$-consistent;

        \item it satisfies the constraints;

        \item we have $\wnego(\lambda) \leq \lambda$.
    \end{itemize}

    Note that the last point does not imply that $\lambda$ is a fixed point of $\wnego$; but by \Cref{lm:if_non_increasing}, it implies $\lambda \geq \lambda^\star$, and then $\pi$ is $\lambda^\star$-consistent, which is enough to conclude.

    As for $\pi$, it is known~\cite[Lemma~5.11]{jacm} that if there is such a play, there is one that has a representation whose size is polynomial in the size of $\Game$, $\lambda$, $\bx$, and $\by$, and that allows us to verify $\lambda$-consistence and constraint satisfaction in polynomial time.

    As for the last point, let us observe that for a requirement $\lambda$ with finite values, we have $\wnego(\lambda) \leq \lambda$ if and only if there exists, for each player $i$, vertex $v \in V_i$ and edge $vw \in E$, a $\lambda$-consistent play $\chi$ from $w$ with $\mu_i(\chi) \leq \lambda(v)$.
    By invoking the same results as above, a certificate for this claim can be guessed and checked in polynomial time.
\end{proof}

\section{The existence of weak SPEs in mean-payoff games} \label{sec:existence}

In this section, we use the weak negotiation function to prove another result: weak SPEs always exist in mean-payoff games.
This result is already known in all prefix-independent games~\cite{existenceWSPE}; however, what follows is a new proof, much shorter than the one cited.
Again, we believe that the techniques used here can be adapted to address some open problem, as we will explain later.

\subsection{The weak negotiation sequence}

Let $\Game$ be a mean-payoff game.
We consider the sequence $(\lambda_n)_n$ of requirements where $\lambda_0$ is the vacuous requirement, and $\lambda_{n+1} = \wnego(\lambda_n)$ for each $n$; we call it the \emph{weak negotiation sequence}.

\begin{remark}
    By induction, for each $n$, we have $\lambda_n \leq \lambda_{n+1}$.
\end{remark}

We can therefore write the following.

\begin{lemma} \label{lm_lambdastar}
    The least fixed point $\lambda^\star$ of the weak negotiation function satisfies $\lambda^\star(v) = \lim_n \lambda_n(v)$ for each $v$.
\end{lemma}

\begin{proof}
    This result is a consequence of Kleene-Tarski's fixed point theorem, which applies here since the weak negotiation function is Scott-continuous, by a proof using the same arguments as the proof of Proposition~3 in~\cite{Concur}.
\end{proof}

An immediate consequence is that a play is a weak SPE outcome if and only if it is $\lambda^\star$-consistent.





\subsection{Good successors, best successors}

For every requirement $\lambda$ and each vertex $v$, we call \emph{good successor} of $v$ with regard to the requirement $\lambda$ every vertex $w$, with $vw \in E$, that maximises the quantity $\inf_{\pi \in \lCons(w)} \mu_i(v\pi)$.
We also define the \emph{best successor} of each vertex $v$ at step $n$ as follows:

\begin{itemize}
    \item the vertex $\bs_0(v)$ is some $w$ such that $vw \in E$;
    
    \item for each $n$, if $\bs_n(v)$ is a good successor of $v$ with regard to the requirement $\lambda_n$, then we maintain $\bs_{n+1}(v) = \bs_n(v)$; otherwise, we define $\bs_{n+1}(v)$ as some good successor of $v$ with regard to the requirement $\lambda_n$.
\end{itemize}

Thus, the vertex $\bs_{n+1}(v)$ is always a good successor of $v$ with regard to the requirement $\lambda_n$, and we change it only when needed to maintain that property.

Finally, for each vertex $v$ and every index $n$, we define the $n$-th \emph{best-successor chain} from $v$ as $\pi^{vn} = \pi^{vn}_0 \pi^{vn}_1 \dots$ where $\pi^{vn}_0 = v$, and $\pi^{vn}_{k+1} = \bs_n(\pi^{vn}_k)$ for each $k$.

\subsection{The best-successor chain is \texorpdfstring{$\lambda_n$}{lambda\_n}-consistent}

We can now prove our key result.

\begin{lemma} \label{lm_pivn}
    For each vertex $v$ and every $n \geq 0$, the best-successor chain $\pi^{vn}$ is $\lambda_n$-consistent.
\end{lemma}

\begin{figure}
    \centering
    \begin{tikzpicture}[thick]
    \tikzstyle{vertexinpath}=[fill=black,minimum size=4pt,inner sep=0pt,circle]
    \tikzstyle{pathFromKp1}=[draw=red]
    \tikzstyle{pathChi}=[draw=green!70!black]
        \node[vertexinpath,label={east:$c_0$}] (v5) at (210:1.5) {}; 
        \node[vertexinpath,label={north:$v$}] (v0) at ($(v5)+(-5,0)$) {}; 
        \node[anchor=east] at (v0.west) {$\pi^{v(n+1)}$:};
        \foreach \x in {1,...,4} {
            \node[vertexinpath] (v\x) at (barycentric cs:v0=5-\x,v5=\x) {};
        }
        \foreach \x in {1,...,5} {
            \pgfmathtruncatemacro{\prev}{\x-1}
            \path[->] (v\prev) edge (v\x);
        }
        \draw (0,0) circle (1.5cm);
        \draw[->] (v5) arc[start angle=210,end angle=190,radius=1.5cm];
        \node[vertexinpath,label={north west:$c_m$}] (cm) at (140:1.5) {};
        \node[vertexinpath,label={$c_k$}] (ck) at (70:1.5) {};
        \node[vertexinpath,label={east:$c_{k+1}$}] (ckp1) at (55:1.5) {};
        \draw[->] (ck) arc[start angle=70,end angle=55,radius=1.5cm];
        \node[vertexinpath,label={north west:$c_\ell$}] (cl) at (-10:1.5) {};
        \node[vertexinpath,label={south:$w$}] (w0) at ($(cl)+(0.75,-0.25)$) {}; 
        \node[vertexinpath] (w3) at ($(w0)+(3,0)+(180:0.75)$) {}; 
        \node[anchor=south,text=red] at ($(w0)+(3,0)+(60:0.75)$) {$\pi^{c_{k+1}n}$}; 
        \foreach \x in {1,...,2} {
            \node[vertexinpath] (w\x) at (barycentric cs:w0=3-\x,w3=\x) {};
        }
        \foreach \x in {1,...,3} {
            \pgfmathtruncatemacro{\prev}{\x-1}
            \path[->,pathFromKp1] (w\prev) edge (w\x);
        }
        \path[->,pathFromKp1] (cl) edge (w0);
        \draw[pathFromKp1] ($(w0)+(3,0)$) circle (0.75cm);
        \draw[->,pathFromKp1] (w3) arc[start angle=180,end angle=150,radius=0.75cm];
        \draw[->,pathFromKp1] (ckp1) arc[start angle=55,end angle=-10,radius=1.5cm];
        \node[vertexinpath] (chi0) at ($(w0)+(0.5,0.75)$) {};
        \node[vertexinpath] (chi1) at ($(chi0)+(0.75,0.5)$) {};
        \node[vertexinpath] (chi2) at ($(chi1)+(1,0.25)$) {};
        \node[vertexinpath] (chi3) at ($(chi2)+(0.75,0.5)$) {};
        \draw[->,pathChi] (w0) edge (chi0);
        \foreach \x in {1,...,3} {
            \pgfmathtruncatemacro{\prev}{\x-1}
            \path[->,pathChi] (chi\prev) edge (chi\x);
        }
        \coordinate (chi4) at ($(chi3) +(0.5,0.5)$);
        \draw[->,pathChi,dash pattern=on 5pt off 2pt on 2pt off 2pt on 2pt off 2pt] (chi3) -- (chi4);
        \node[anchor=south west,text=green!70!black] at (chi4) {$\chi$};
    \end{tikzpicture}
    \caption{Play $\pi^{v(n+1)}$ and deviations from it (used in the proof of Lemma~\ref{lm_pivn}).}
    \label{fig:bestSuccChainConsistent}
\end{figure}
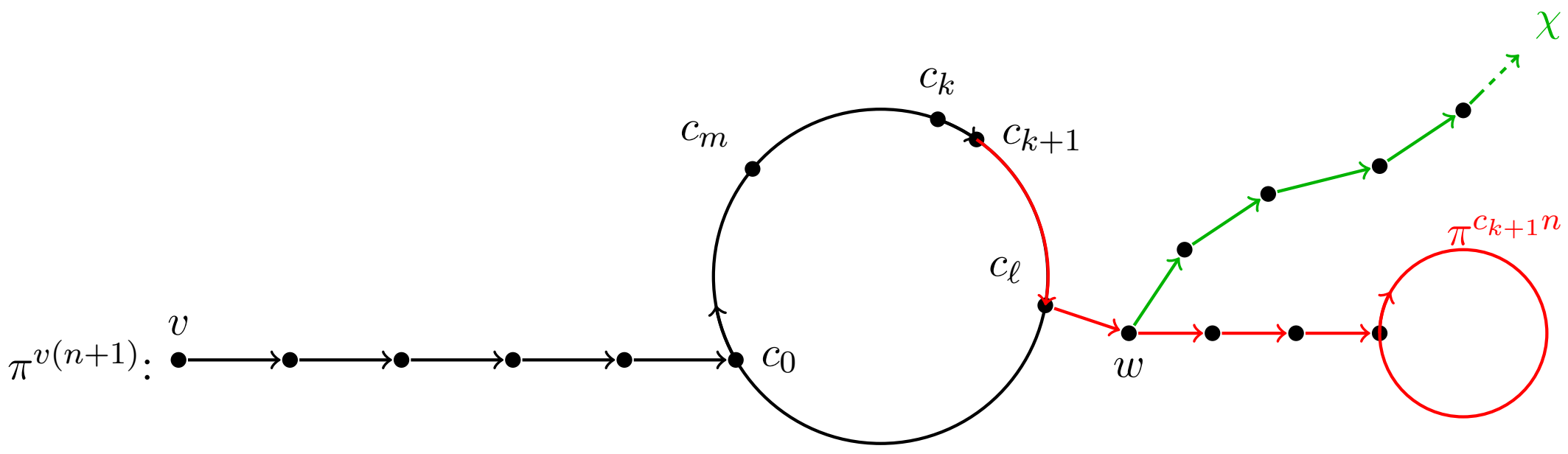

\begin{proof}
    We proceed by induction on $n$: for $n = 0$, the property is true, because every play is $\lambda_0$-consistent.
    Let us now assume toward contradiction that for some $n \geq 0$, each $\pi^{vn}$ is $\lambda_n$-consistent (the induction hypothesis), but that there exists $v$ such that $\pi^{v(n+1)}$ is not $\lambda_{n+1}$-consistent.
    Let us simply write $\pi = \pi^{v(n+1)}$.
    A visual illustration of this proof is given by \Cref{fig:bestSuccChainConsistent}.

    Since $\pi$ is a play in which each vertex is always followed by the same successor, it has the form $\pi = hc^\omega$, where $h = h_0 \dots h_{p-1}$ is a simple history, and $c = c_0 \dots c_{q-1}$ is a simple cycle.

    \begin{claim}
        The play $c^\omega$ is not $\lambda_{n+1}$-consistent.
    \end{claim}

    \begin{proof}
        Let us assume toward contradiction that there is a last index $k < p$ such that $\mu_i(\pi_{\geq k}) < \lambda_{n+1}(\pi_k)$ (where $i$ is the player controlling $\pi_k$).
        Then the play $\pi_{> k}$ is $\lambda_{n+1}$-consistent, and therefore $\lambda_n$-consistent.
        
        But since $\pi_{k+1} = \bs_{n+1}(\pi_k)$, the vertex $\pi_{k+1}$ is a good successor of $\pi_k$ with regard to the requirement $\lambda_n$, i.e., every $\lambda_n$-consistent play starting from the vertex $\pi_{k+1}$ gives player $i$ a payoff at least $\lambda_{n+1}(\pi_k)$.
        In particular, that is the case of the play $\pi_{\geq k+1}$, which is a contradiction.

        There is therefore no $\lambda_{n+1}$-consistent suffix of $\pi$, and in particular $c^\omega$ is not.
    \end{proof}

    Therefore, there exists an index $k < q$ such that $\mu_i((c_k \dots c_{q-1} c_0 \dots c_{k-1})^\omega) < \lambda_{n+1}(c_k)$, where $i$ is the player controlling $c_k$.
    The vertex $c_{k+1}$ (if $k = q-1$, we set $c_q = c_0$) is a good successor of $c_k$ with regard to the requirement $\lambda_n$, hence every $\lambda_n$-consistent play from $c_{k+1}$ gives player $i$ a payoff greater than or equal to $\lambda_{n+1}(c_k)$.
    It is in particular the case of the best-successor chain $\pi^{c_{k+1}n}$, which is $\lambda_n$-consistent by induction hypothesis.
    As a consequence, the plays $\pi^{c_{k+1}n}$ and $(c_{k+1} \dots c_{q-1} c_0 \dots c_k)^\omega$ are different.

    Therefore, there exists an index $\l$ such that the best successor of $c_\l$ has changed between step $n$ and step $n+1$: let us write $w = \bs_n(c_\l) = \pi^{c_{k+1}n}_{(\l-k)} \neq c_{\l+1} = \bs_{n+1}(c_\l)$.
    Then, by definition of best successors, the vertex $w$ (which, if $n \geq 1$, was a best successor of $c_\l$ with regard to $\lambda_{n-1}$) is no longer a good successor of $c_\l$ with regard to the requirement $\lambda_n$: this means that there exists a $\lambda_n$-consistent play $\chi$ from $w$ such that $\mu_j(\chi) < \lambda_{n+1}(c_\l)$, where $j$ is the player controlling $c_\l$.
    And since $c_{\l+1}$, on the other hand, is a good successor of $c_\l$ with regard to $\lambda_n$, we obtain that for every $\lambda_n$-consistent play $\xi$ from $c_{\l+1}$, we have $\mu_j(\xi) > \mu_j(\chi)$.

    But then, let us consider the play $\xi = c_{\l+1} \dots c_{q-1} c_0 \dots c_\l \chi$: if that play was $\lambda_n$-consistent, we would then have $\mu_j(\xi) > \mu_j(\chi)$ by the above, whereas prefix-independence gives $\mu_j(\xi) = \mu_j(\chi)$.
    Therefore, that play is not $\lambda_n$-consistent.
    However, its suffix $\chi$ is.
    Therefore, there exists a last index $m \in \{\l+1, \dots, q-1, 0, \dots, \l\}$ such that $c_{m+1} \dots c_\l \chi$ is $\lambda_n$-consistent, but $c_m \dots c_\l \chi$ is not.
    Then, if $j'$ is the player controlling $c_m$, we have $\mu_{j'}(\chi) < \lambda_n(c_m)$.

    Let us separate two cases, and first assume $m \neq \l$.
    Then, the vertex $c_{m+1} = \bs_{n+1}(c_m)$ is a good successor of $c_m$ with regard to $\lambda_n$, which implies that every $\lambda_n$-consistent play from $c_{m+1}$ gives player $j'$ at least the payoff $\lambda_{n+1}(c_m)$; and it is in particular the case of the play $c_{m+1} \dots c_\l \chi$.
    Therefore, we have $\mu_{j'}(c_{m+1} \dots c_\l \chi) = \mu_{j'}(\chi) \geq \lambda_{n+1}(c_m) \geq \lambda_n(c_m)$: contradiction.

    Similarly, if $m = \l$, we have $\mu_j(\chi) \geq \lambda_n(c_\l)$: this is clear when $n = 0$, and a consequence of $w = \bs_n(c_\l)$ when $n \geq 1$.
    And it is a contradiction.
\end{proof}

\subsection{Conclusion}

We can now conclude on the following theorem.

\begin{theorem}
    In every mean-payoff game, there exists at least one weak SPE.
\end{theorem}

\begin{proof}
    Let us first notice that, by Theorem~\ref{thm:wnego} and Lemma~\ref{lm_lambdastar}, this theorem is equivalent to the following: for every mean-payoff game $\Game$ and every vertex $v$, there exists a $\lambda^\star$-consistent play from $v$.
    That is actually equivalent to say that $\lambda^\star(v) \neq +\infty$, because if $\lambda^\star(v) = +\infty$, there is clearly no $\lambda^\star$-consistent play from $v$; and conversely, if $\lambda^\star(v) \neq +\infty$, then we can choose a good successor $w$ of $v$ with requirement $\lambda^\star$, which will then satisfy $\inf_{\pi \in \lambda^\star\Cons(w)} \mu_i(\pi) = \wnego(\lambda^\star)(v) = \lambda^\star(v) \neq +\infty$, which implies that there exists a $\lambda^\star$-consistent play $\pi$ from $w$ and that the play $v\pi$ is actually $\lambda^\star$-consistent from $v$.

    Let us now assume that $\lambda^\star(v) = +\infty$: then, since mean-payoff functions are bounded, we have $\lambda_n(v) = +\infty$ for some $n \in \NN$.
    But that is impossible since by Lemma~\ref{lm_pivn}, the play $\pi^{vn}$ is $\lambda_n$-consistent.
\end{proof}

\subsection{An open problem}

To the best of our knowledge, the proof presented above is the first existence proof based on (a variant of) the negotiation function.
This suggests that variants of the negotiation function could be used again in the future to prove similar results, including long-standing conjectures about equilibria existence.

One candidate is the open question raised by Flesch, Kuipers, Schoenmakers, and Vrieze \cite{DBLP:journals/ijgt/KuipersFSV16}: assuming that the players are allowed to randomise their strategies, is it the case that in every mean-payoff game, for every $\epsilon > 0$, there exists an $\epsilon$-SPE?
It is known that SPEs may not exist, even in the randomised case~\cite{SOLAN2003911}.
But for $\epsilon$-SPEs, the known examples suggest that only a mild form of randomisation is necessary, with some action being performed with an infinitesimal probability, whose purpose is precisely to punish infinite deviations.
This dynamic could be captured with a new variant of the negotiation function, and a proof similar to ours may then positively close this question.
\section{Discussion}\label{sec:conclusion}

After a quick review of the results already obtained from the standard negotiation function, we have introduced the \emph{weak negotiation function}, a variant that captures weak SPEs in the same way the negotiation function captures SPEs.
This characterisation of weak SPEs gives a new computational result: the constrained existence problem of weak SPEs is $\NP$-complete in mean-payoff games, matching the complexity of the analogous problem in parity games~\cite{DBLP:journals/iandc/BrihayeBGR21}, or with SPEs~\cite{jacm}.

We have also given a new proof of a known result: 
weak SPEs are guaranteed to exist in every mean-payoff game.
This proof is significantly more succinct than the known one, suggesting that the negotiation function and its variants still provide a strong toolbox to address equilibria problems in graph games.

\acknowledgements{Part of this work was done when I was FNRS aspirant at the Université libre de Bruxelles, and another one as part of the project VAMOS that has received funding from the European Research Council (ERC), grant agreement No 101020093.
I warmly thank Marie van den Bogaard and Jean-François Raskin for their kind and rigorous supervision during my PhD, during which these results have been produced.
I also thank Mathieu Sassolas, for fruitful discussions that have led to some of those results, and for carefully letting me have my naps when we were sharing the same office, which played a crucial role in this work.}

\bibliographystyle{fundam}
\bibliography{bibli}

\appendix

\end{document}